# A microsimulation of spatial inequality in energy access: A Bayesian multi-level modelling approach for urban India

**André Paul Neto-Bradley [a] \*, Ruchi Choudhary [a,b], Peter Challenor [b,c]**

**a** Energy Efficient Cities Initiative, Department of Engineering, University of Cambridge, Trumpington Street, CB1 2PZ, UK

**b** The Alan Turing Institute, British Library, 96 Euston Road, London NW1 2DB, UK

**c** Department of Mathematics, College of Engineering, Mathematics and Physical Sciences, University of Exeter, Exeter EX4 4QE, UK

**\* Corresponding author:** André Paul Neto-Bradley, Energy Efficient Cities Initiative, Department of Engineering, University of Cambridge, Cambridge CB1 2PZ, U.K. Email: apn30@cam.ac.uk,

## Abstract

Access to sustained clean cooking in India is essential to addressing the health burden of indoor air pollution from biomass fuels, but spatial inequality in cities can adversely affect uptake and effectiveness of policies amongst low-income households. Limited data exists on the spatial distribution of energy use in Indian cities, particularly amongst low-income households, and most quantitative studies focus primarily on the effect of economic determinants. A microsimulation approach is proposed, using publicly available data and a Bayesian multi-level model to account for effects of current cooking practices, local socio-cultural context, and spatial effects. This approach offers previously unavailable insight into the spatial distribution of fuel use and residential energy transition within Indian cities. Uncertainty in the modelled effects is propagated through to fuel use estimates. The model is applied to four cities in the south Indian states of Kerala and Tamil Nadu, and comparison against ward-level survey data shows consistency with the model estimates. Ward-level effects exemplify how wards compare to the city average and to other urban area in the state, which can help stakeholders design and implement clean cooking interventions tailored to the needs of households.

## Keywords

# Introduction

## *Clean Cooking and Energy Access*

Use of traditional solid biomass fuels for cooking is a reality for many Indians, with just under 50% of the population still using solid biomass fuel to meet some part of their cooking needs (International Energy Agency, 2020a). This is the case despite approximately 95% of the population having access to distribution of Liquified Petroleum Gas (LPG) following the recent Pradhan Mantri Ujjwala Yojana (PMUY) programme (International Energy Agency, 2020b). Even in urban areas biomass use persists and is most prevalent amongst low-income households (Ahmad and Puppim de Oliveira, 2015), who often face spatial inequality in their access to utilities (Bhan and Jana, 2015). Use of solid biomass fuels in cooking stoves releases pollutants including carbon monoxide and particulate matter in the form of soot, that contribute to the over 600,000 annual deaths attributed to indoor air pollution in India (Pandey *et al.*, 2021).

Recent studies have shown that urban households can follow different pathways to adopting clean cooking based on the local interaction of socio-economic features, community, behaviours, and infrastructure, that require tailored and targeted intervention (Neto-Bradley, Choudhary and Bazaz, 2020). Targeting efforts to address different sets of barriers at an urban scale is complicated by the lack of available data and methods suited to a developing country context. Most quantitative studies on residential energy use in India use regressions based on expenditure and income levels (Farsi, Filippini and Pachauri, 2007; Ekholm *et al.*, 2010), asset ownership and socio-economic variables such as education (Ahmad and Puppim de Oliveira, 2015; Sankhyayan and Dasgupta, 2019), or type of employment (Kemmler, 2007; Sehjpal *et al.*, 2014) to predict the effect on energy consumption. These studies assume all households will respond in the same way to economic stimuli. A further drawback of such models is that interpretation of the model outputs as they relate to local context may not be intuitive and can require additional learning, limiting their real world usefulness (Nochta *et al.*, 2020).

On the other hand more localised studies characterising the role of energy related practices and socio-cultural, economic, and technical factors using qualitative approaches in the Global South have shown the importance of household practices in understanding energy transition (e.g. Bisaga and Parikh (2018)). Mixed method approaches which attempt to meaningfully integrate quantitative and qualitative methods have been limited, but several recent studies have adopted such an approach to shed light on the circumstances of low-income urban households in India (Khosla, Sircar and Bhardwaj, 2019; Neto-Bradley *et al.*, 2021). A key takeaway from these studies is that considering individual household-level decisions, behaviours, and context is key to identifying opportunities to facilitate uptake of clean energy among low-income urban households. Doing so at scales sufficiently large to understand energy behaviours across entire cities remains a challenge.

## *Microsimulation and bottom-up models*

Microsimulations are a form of bottom-up model that simulate features and actions of synthetic representative individuals in a population (Frayssinet *et al.*, 2018). Such methods have been used in transport research and epidemiology, for example to investigate cardiovascular disease risks throughout a national population (Knight *et al.*, 2017). More recently such methods have been used in urban-scale energy studies with the use of simulated building stocks where the individuals are buildings rather than people (Booth, Choudhary and Spiegelhalter, 2012; Zakhary *et al.*, 2020).

Due to reasons of privacy and cost, individual-level representative datasets at urban or district scale are often not available (Casati *et al.*, 2015). Microsimulation studies generate synthetic populations in the absence of actual individual level data using available public aggregate datasets, either using a single dataset if all required information is contained or using a combination of datasets (Barthelemy and Toint, 2013). Such approaches have recently been used at a city scale to determine CO2 emission density (Tirumalachetty, Kockelman and Nichols, 2013), and household gas and electricity use in US cities (Zhang *et al.*, 2018). Generating a population of individuals enables simulation of outputs based on individual level features and behaviours, better capturing heterogeneity and inequalities across the area of study.

*Uncertainty in Modelling*

Microsimulation modelling naturally embodies uncertainties in the estimation of the synthetic population. These are challenging to quantify methodologically. The absence of test data generally renders any form of uncertainty management of the model parameters impossible. However, it is feasible to include and propagate the uncertainties arising from the variation across households in the dataset used for generating the synthetic population. As such, these represent the first order uncertainties, or the heterogeneity across similar clusters of households. The advantage of including these in the analysis is twofold: First, it helps understand the variability of outputs across different population groups; and second, it enables calibration of parameters as new data becomes available.

This paper presents a microsimulation approach which uses a Bayesian multi-level model to estimate cooking fuel consumption across city wards, and thus provide a disaggregated urban scale view of spatial inequality in clean cooking access with quantification of uncertainties. This approach is based on the expectation that household-level energy use is conditioned by household habits and practices, local socio-economic and cultural context, and spatial effects. The novelty of this study lies both in its use of a microsimulation approach to model clean cooking access at a household-level, and the systematic treatment of uncertainty in estimating fuel consumption in a Global South context.

The approach combines publicly available data from the census and nationally representative surveys, to generate a synthetic population of individual households. Markov Chain Monte Carlo (MCMC) sampling is used to estimate parameters for a multi-level model, which predicts fuel use and fuel stacking prevalence (use of multiple fuels i.e. both biomass and LPG) at a household scale while accounting for group effects of cooking fuel choice, and spatial and non-spatial heterogeneity. This model is applied to four different case study cities in southern India, namely Coimbatore, Tiruchirappalli, Kochi, and Trivandrum, details of the study cities are summarised in Supplementary Figure S.1, with an accompanying rationale for selection. Primary data collected from selected wards in these cities is compared to model outcomes for a consistency check. The model outputs enable the identification of those wards within a city that are worst affected by continued solid biomass fuel, and offer insight into how wards compare to each other, as well as to the average urban household in the state. With a new Indian census being conducted this year this approach will provide a means to update and track the state of clean cooking transition in urban India in the coming decade. The remainder of this paper will introduce the methodology, before assessing model performance and discussing model outputs. The paper concludes with a discussion of the features, utility, and policy relevance of this approach.

# Methodology

*Overview*

The aim of this approach is to understand differences across city wards by modelling household-level energy use, but household-level data only exists per state and district, while available ward-level data includes tallies for a limited set of socio-economic variables and does not provide details on fuel consumption. As a result of this discrepancy between desired scale of analysis and availability of public data multiple data sources at differing geographic scales are combined to model the spatial inequality in access to clean cooking across a city.

Cooking fuel use and energy choices at a household-level are estimated taking into account household practices, locally acting socio-economic and community factors, and spatial effects. This requires generating a representative population of households whose combination of socio-economic characteristics is known. Iterative proportional fitting (IPF) is used to first generate a synthetic population of households, representative at the ward-level. Each of these synthetic households are assigned to a primary fuel choice group using a categorical logistic regression based on socio-economic predictors. The fuel consumption and fuel stacking prevalence are then estimated for each household using a multi-level model with model coefficients varying on the basis of primary cooking fuel choice of the household and taking into account ward-level effects. Figure 1 provides a schematic overview of the model structure.

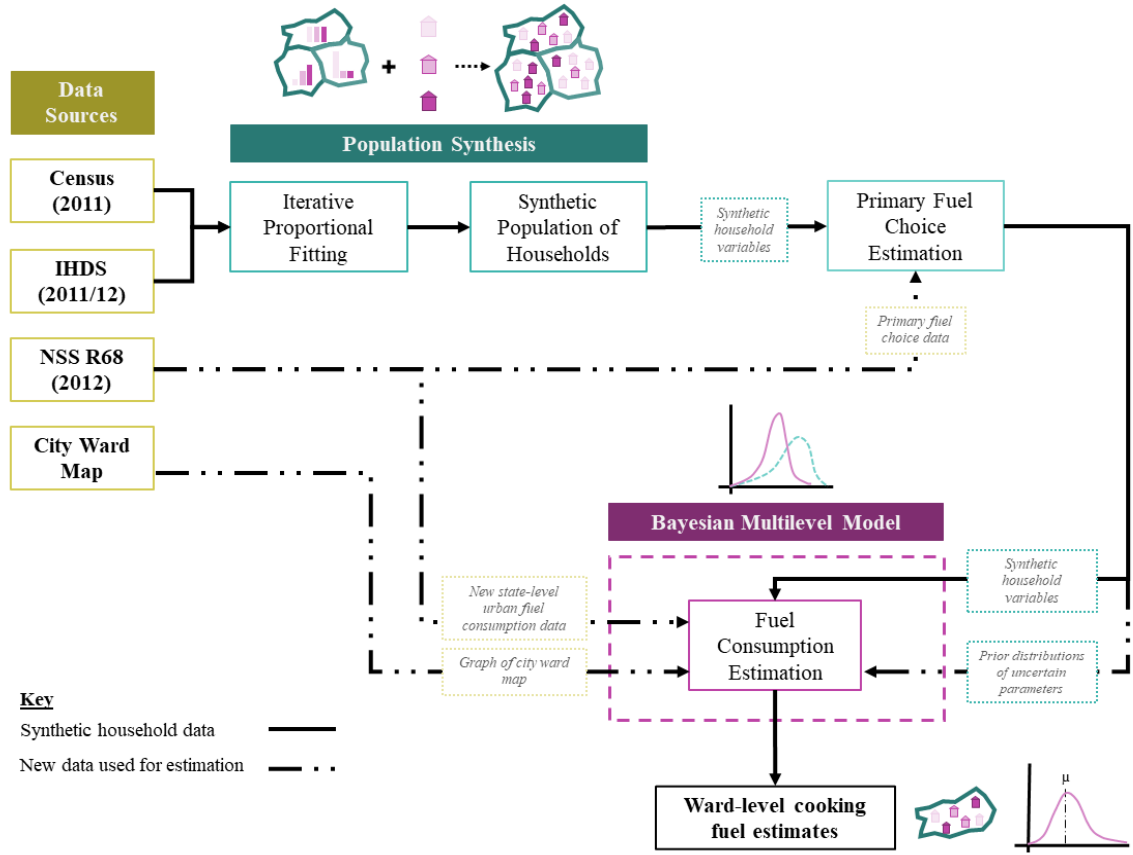

Figure 1: Overview of microsimulation approach including data sources and outputs.

## Data

The household-level data from the Indian Human Development Survey 2011 (IHDS-II) (Desai and Vanneman, 2015) and the 2011 census (Government of India, 2011) ward-level tables of household asset ownership are used to generate the synthetic population. Maps of the Census 2011 ward boundaries, and National Sample Survey consumer survey data from Round 68 (National Sample Survey Organisation, 2013) are used as inputs for the multi-level model. The performance of the model is examined using survey data from 24 low-income wards across the four cities. Further details of the data sources including sample characteristics of ward-level survey data are detailed in the Data section in the Supplementary Material.

## Household population synthesis

A synthetic population is generated using IPF, a method also known as matrix raking, as introduced by (Deming and Stephan, 1940). This approach does have some drawbacks including reliance on categorical data, and ability to only control for household or individual level attributes (Casati *et al.*, 2015). However this is well suited for this model given the data publicly available on Indian households, and that the household is the smallest unit of analysis not the individual. IPF offers the benefit of being computationally efficient, simple to use (including practical guidance on performing IPF in R (Lovelace and Dumont, 2016)), and converges to a single solution (Fienberg, 1970). The IHDS household-level data is used as the microdata consisting of a sample of representative individual urban households from the state. The census ward-level data is used as a contingency table of constraints to generate weightings for the individuals in the microdata. A detailed explanation of how IPF uses this data to generate a synthetic population, what assumptions are made, and how weightings are integerised is detailed in Figure S.2 and Table S.2 in the Supplementary Material.

Synthetic households are assigned to a primary cooking fuel group representative of current cooking practices. A categorical logistic regression of the form shown in equation 1 is used to infer the primary cooking fuel $j$ of a household $i$ based on the socio-economic characteristics of the household. The predictors used are household expenditure, majority religion membership, ration card possession, caste category, and income frequency. The regression coefficients are estimated from the NSS data for the respective state. The household is assigned to one of six primary cooking fuel choice groups: No cooking, Biomass, Kerosene, Low LPG (uses less than a cylinder every month), High LPG (uses more than a cylinder every month), or Electricity.

$$j[i] = \beta_0 + \beta_1 exp_i + \beta_2 relig_i + \beta_3 rationcard_i + \beta_4 caste_i + \beta_5 incfreq_i \qquad \text{Eq. 1}$$

Generation and validation of the synthetic population for each of the four cities was done using base packages in R as well as the 'ipfp' package (Blocker, 2016).

*Multi-level model specification*

A multi-level model is used to estimate the magnitude of biomass and LPG use for each synthetic household on the basis of its socio-economic features and location. The model has three components which aim to address intrinsic features and practices of the household, extrinsic local socio-economic and cultural features, and spatial effects. The household-level component of the model uses an idealised linear relationship for the household's mean fuel use with expenditure and household size (no. of persons) as predictors. These represent the intrinsic component of cooking fuel use, dependant on the features of the household. Coefficients for the household-level predictors are allowed to vary by primary cooking fuel group. This is a non-spatial, and non-random group effect.

A novelty of this study is to quantify the influence of local socio-economic and cultural features on fuel use and access. These are extrinsic features which relate to the household's interaction with its wider neighbourhood and community and spatial effects from its location within the city. Quantifying the impact of local features is complicated by the unmeasurable nature of some socio-cultural determinants of clean cooking access. This is addressed by adapting the approach of the Besag-York-Mollie model (Besag, York and Mollié, 1991) used in spatial epidemiology studies (DiMaggio, 2015) which uses an Intrinsic Conditional Auto-Regressive (ICAR) to capture spatial effects and a random effects coefficient to capture extrinsic features. Similar approaches have been used in district-wise building energy use models (Choudhary and Tian, 2014), and recent studies have explored improved approaches for implementation (Morris *et al.*, 2019).

The statistical model is shown in equation 2, where for a household $i$ in cooking fuel group $j$ and city ward $w$ fuel use is assumed to follow a normal distribution with a mean given by the sum of the household and ward-level components and with a precision $\sigma$. In equation 2 $\phi_{w[i]}$ is the ICAR component and $\theta_{w[i]}$ is the random effects coefficient.

$$fuel_{[i]} \sim N\big(a_{0\,j[i]} + a_{1\,j[i]} exp_{[i]} + a_{2\,j[i]} size_{[i]} + \phi_{w[i]} + \theta_{w[i]}, \sigma\big) \qquad \text{Eq. 2}$$

Household-level predictor coefficients $a_0, a_1,$ and $a_2$ are defined by the following parameters:

$$a_{0[i]} \sim N(\mu_{a0}, \sigma_{a0})$$

$$a_{1[i]} \sim N(\mu_{a1}, \sigma_{a1})$$

$$a_{2[i]} \sim N(\mu_{a2}, \sigma_{a2})$$

Parameters for the distributions of coefficients $a_0$, $a_1$, and $a_2$ are estimated from the NSS consumer survey data. A prior is set on the parameters using the fuel use values embedded in the synthetic population, which have been derived from the IHDS micro dataset. Four different versions of the model were implemented to compare the performance of the different components of the model. These four variants are a Basic Model which includes only household-level effects; a Random Effects Model which includes household-level predictors and the random effect coefficient; an ICAR Model which includes household-level predictors and an ICAR component; and a combined model which includes all three components (shown in equation 2). Details of model variant formulations, priors, and parameter definitions are detailed under Multi-level Model Variants in the Supplementary Material.

Models for each city are estimated separately, using NSS consumer survey data for the respective state. All expenditure and fuel consumption values are normalised and a square root transformation is also applied to fuel consumption values. Estimation of the parameters of the model is done using Stan which performs full Bayesian inference using Hamiltonian Monte Carlo. Stan's No U-Turn Sampling (NUTS) performs better than alternative Gibbs or Metropolis algorithms for models with complex posteriors (Homan and Gelman, 2014). The RStan interface enables input and output data handling in R. It is important to note that as Gelman (2006) points out, multilevel models are advantageous for making predictions as they can estimate the effect of individual predictors as well as the group-level mean, but these cannot be necessarily interpreted as causal and this should be kept in mind when examining outputs.

## Model Performance

By comparing the model outputs to the survey data available for selected wards, the consistency of the model and its four variants can be examined. Figure 2 compares the model estimates for LPG and firewood use in each of the 24 wards included in the survey dataset. The points indicate the mean and lines show the 95% confidence interval of each model and the survey data ward-by-ward. The wards are ordered by their surveyed mean fuel consumption. Figure 2 shows that with a few exceptions the survey means fall within the confidence interval for the combined model estimates, and in a majority of cases those of the random effects model too. Notice that the combined model estimates broadly follow the variation seen in the survey means with wards with lower survey means likely to have a lower model estimate than those with higher survey means. This is noteworthy as these wards were selected for the survey because of their low socio-economic status and greater likelihood to have lower access to LPG and use greater amounts of firewood, thus they do not represent the full spectrum of wards but rather the lower end of LPG and upper end of firewood consumption.

Comparing the different model variants, the basic model has the narrowest confidence intervals, and the interval increases for the random effects and ICAR models, with the combined model featuring the widest interval. This reflects the added uncertainty the estimates from these models account for. Recall that the basic model infers coefficients for predictors from NSS data on all urban households in the state, including those not in these cities. It represents the estimate of fuel use for that ward's population of households assuming they are in an 'average' city in that state. The random effects model includes a coefficient that represents the impact of the 'city effect' versus the average city in the state. The ICAR model represents how the ward compares to the wards around it. Accounting for each of these effects adds uncertainty to the estimate and so the combined model has the widest confidence interval.

Model variant estimates differ in performance when compared to the survey data. For firewood consumption there is little difference in the mean estimates between the four model variants, where the survey data means fall within the confidence interval of model outputs for most wards, particularly those with lower consumption levels. This indicates that the levels of consumption in each ward are similar to those in the average urban area in the state and that local spatial effects have a relatively modest effect on magnitude of consumption. The discrepancy between model estimates and survey data for some wards in Tiruchirappalli (abbreviated to Trichy in the figure) could represent a transient random effect, for example added use of fuel in the month surveyed due to a festival or greater availability of fuel.

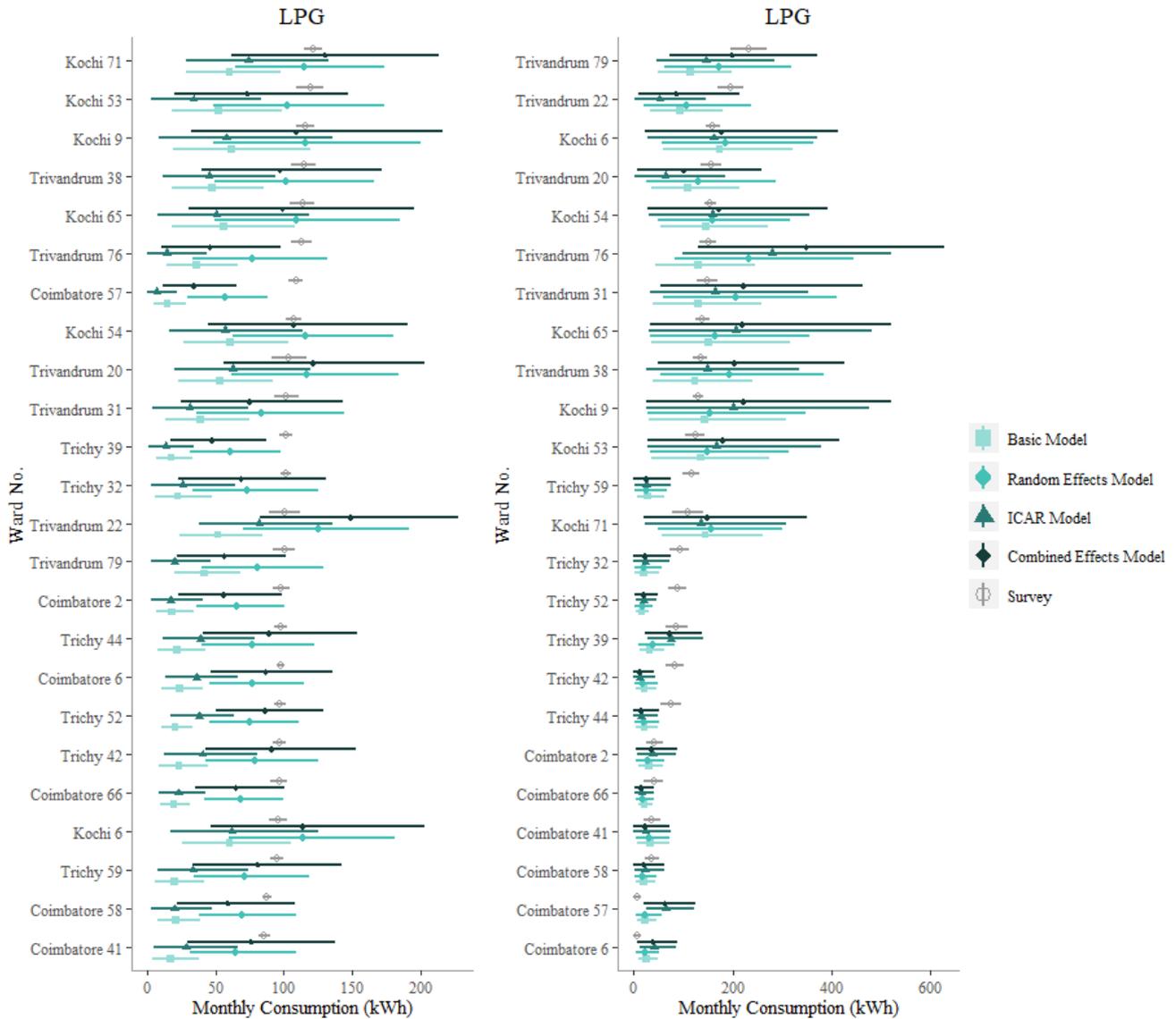

Figure 2: Comparison of model outputs and survey data for selected wards. Point values represent mean values from the model or survey samples, and lines indicate the 95% confidence interval.

The situation with LPG is rather different, where the basic model underestimates LPG use and the random effects and combined models are more compatible with the survey data. In this instance the random effect coefficient provides the bulk of the adjustment from the basic model with the ICAR component adding only small additional adjustments. It is worth noting two points here; Firstly, the fact that these cities are larger than the average urban centre within their states would suggest that they are likely to have a better network of LPG suppliers than the average city. In addition, the survey data was collected in 2018-19 coinciding with the government's PMUY LPG connection initiative which targeted certain groups of low-income households, which possibly explains why in wards such as Coimbatore's 57th or Trivandrum's 76th the models underestimate LPG consumption.

## Results & Discussion

The focus of this microsimulation approach was to estimate spatial inequality in access and use of clean cooking fuels across cities, with the expectation that variation at a ward-level arises due to household context, local socio-economic and cultural factors, and spatial effects. Examining the various components of the model is instructive to understand their influence on outputs and how uncertainty is propagated. At the heart of the basic model is a multi-level normal linear model whose coefficients vary according to the primary cooking fuel group a household is predicted to belong to. Figure 3 shows the estimated linear coefficients of the LPG consumption model for Tiruchirappalli across the six different primary cooking fuel groups. The prior is represented by the dotted line. Due to absence of a comparable primary cooking group variable in the raw synthetic population data the prior does not vary by group. Despite this limitation, using the synthetic population data to form a prior ensures consistency and makes for a reasonable prior.

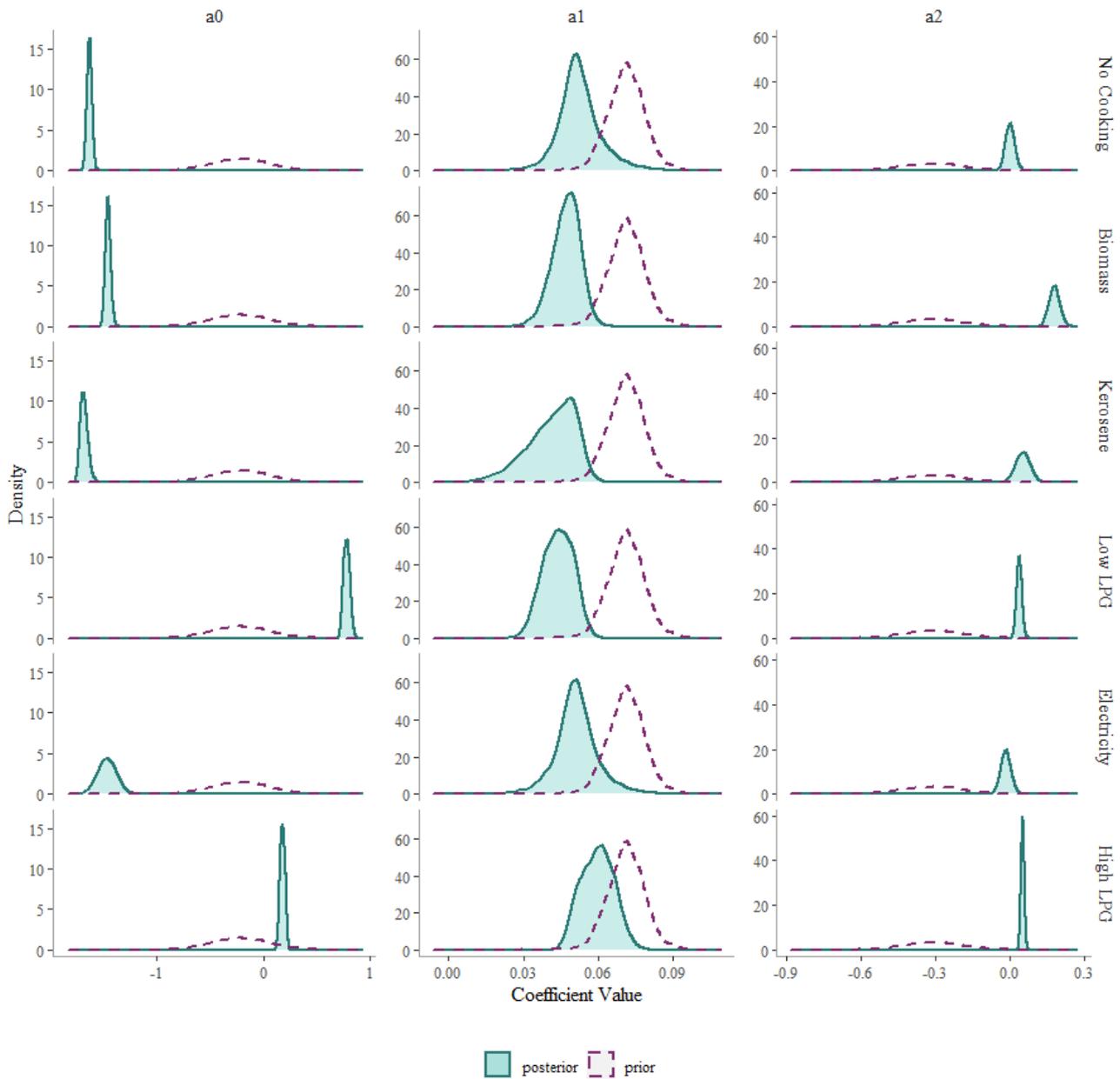

Figure 3: Linear coefficients of LPG household model by primary cooking fuel group for the city of Tiruchirappalli.

The new data from the NSS consumer survey helps reduce uncertainty in the model coefficients with posteriors having smaller variances. While all groups have the same prior, the new data and the model design allowing coefficients to vary by primary cooking fuel group greatly reduce uncertainty. For groups using kerosene and electricity as primary cooking fuels, some coefficient posteriors have greater variance than other groups. This is because in the case of Tiruchirappalli the NSS consumer survey data had considerably fewer instances of households in these two groups and there was variability in fuel use amongst these few cases. This resulted in posterior distributions for coefficients with slightly greater variance. Expenditure slope coefficient a2 for these groups shows uncertainty with respect to the positive or negative nature of this relationship, covering a range of values from approximately -0.05 to 0.15 for the kerosene group, and -0.1 to 0.1 for the electricity group. Interestingly the coefficients for household size a1, while having slightly different distributions between groups, have similar values within a range of 0.04 to 0.08, suggesting the effect of this predictor does not vary much with primary cooking fuel choice.

The components of the model that account for local socio-cultural features and spatial effects are the random effects and ICAR coefficients. These location-related coefficients offer valuable insight for the targeting of interventions as they quantify the impact of local socio-economic and community factors relative to the average ward in the state and in that city. The random effects coefficients across the city as a whole indicates whether the fuel consumption of that city is above, below, or about average for the state. Figure 4 shows the distribution of estimated random effect coefficients across all wards in each of the four cities for both firewood and LPG consumption. They indicate all cities are highly likely to have a positive random effect coefficient for LPG with some small difference in magnitude between them, nonetheless, indicating LPG consumption above state urban area average. The same is not true for firewood consumption where, with the exception of Trivandrum, the random effect coefficient value for firewood consumption across the cities is normally distributed close to zero indicating that firewood consumption levels are close to the state urban average. Note the variance in the coefficient does vary slightly between cities and fuels, reflecting differences in uncertainty in the magnitude of difference between local fuel consumption compared to the state-wide average urban household.

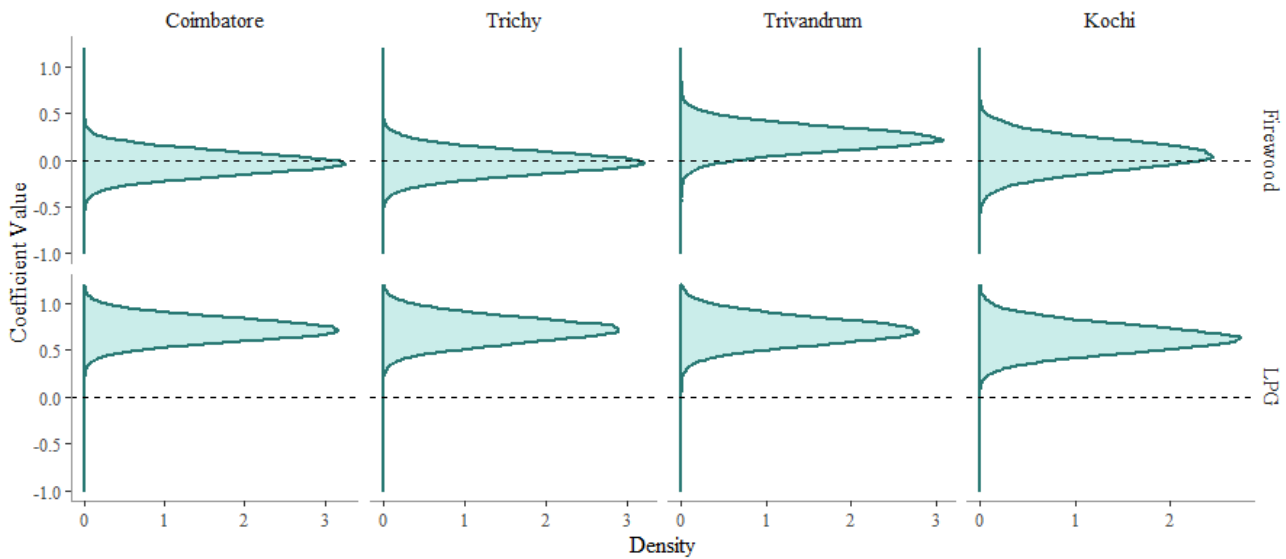

Figure 4: Distribution of random effects coefficients for both firewood and LPG across cities.

Figure 5 shows the mean ward-level random effects and spatial smoothing coefficients for LPG consumption across the city of Tiruchirappalli. The mean random effect coefficient shows negligible variation between wards and indicates that the city as a whole has a level of LPG use well above that of the average urban area in Tamil Nadu. Meanwhile the ICAR coefficient shows how city wards compare to each other. The central southern wards of the city have a positive coefficient indicating higher LPG use than the average ward, while the eastern wards underperform relative to the average ward. Analysis of the spatial distribution of ICAR coefficients provides a clear indication of which wards are below average in terms of uptake of LPG and may thus need additional targeted intervention.

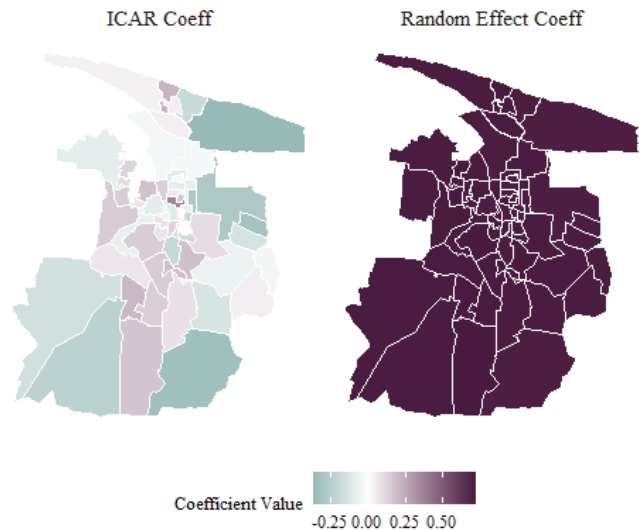

Figure 5: Ward-level random effects and ICAR coefficient values for Tiruchirappalli.

Putting these model components together they add up to produce the model fuel consumption estimates as shown in Figure 6, which illustrates the distribution of estimates of monthly LPG consumption for each of the four model variants in Ward 38 of Tiruchirappalli. The ICAR model indicates that ward 38 is likely to have above average LPG consumption compared to the city average, and the random effects model indicates that Tiruchirappalli's LPG consumption is also likely to be above the state average. The contribution of these two coefficients are added to the basic model estimate in the combined model, resulting in a boost to the mean estimate. The uncertainty from the various individual components adds up with the ICAR and random effects models having greater variance than the basic model and this variance compounds in the combined model which accounts for the uncertainty in all three constituent model components.

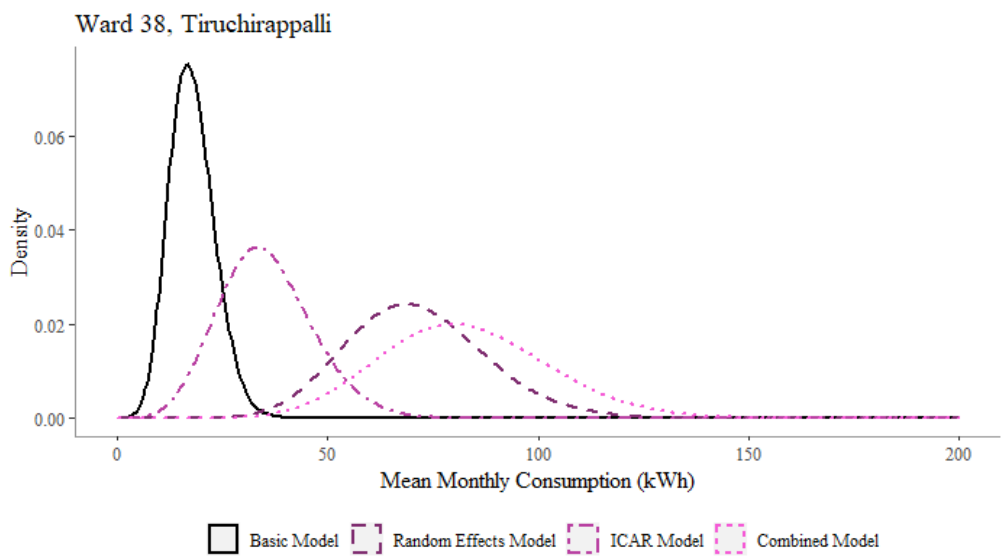

Figure 6: Distribution of LPG Consumption estimates for each of four model variants for Ward 38 in Tiruchirappalli.

The impact of the different models is illustrated spatially in Figure 7, which shows the LPG consumption estimates across city wards for each of the four models. The estimates from the basic model show only small variation between wards while the random effects model boosts the estimates of fuel consumption relative to the basic model, although still preserving the relative homogeneity between wards. The spatial smoothing model introduces greater variation between wards with either an increased or decreased mean estimate compared to the base model. The combined model features both the city wide increase in mean estimated consumption while also including greater variation between wards, more fully capturing the spatial variability.

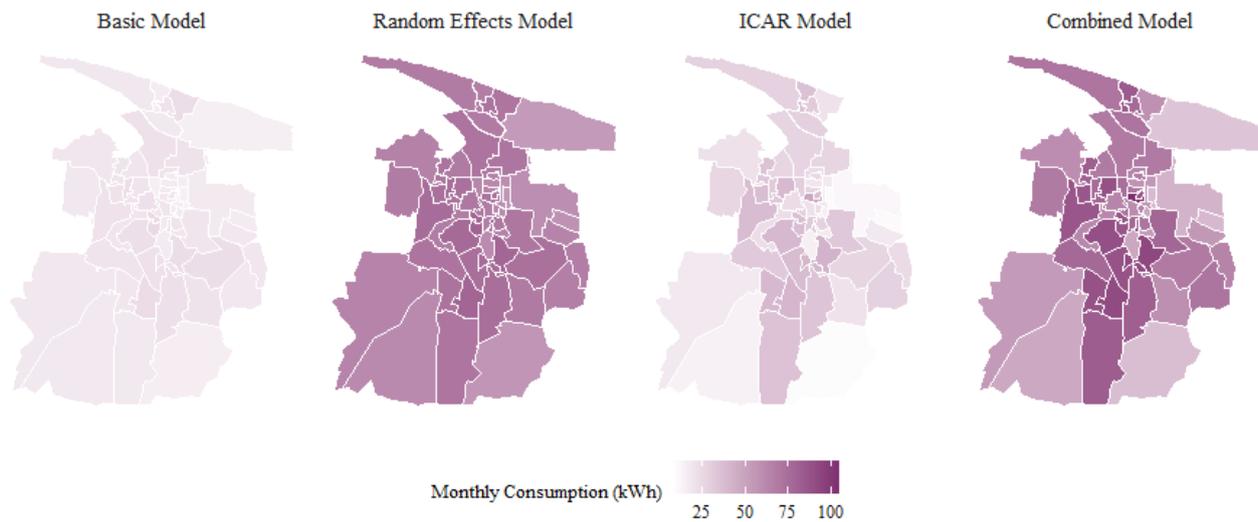

Figure 7: Comparison of model variant estimates for mean monthly LPG consumption in Tiruchirappalli mapped by city wards.

This approach can provide valuable insight for policy makers and stakeholders by locating wards most likely to have low sustained and exclusive clean cooking fuel use. The representative population of synthetic households offers an additional benefit by providing insight into ward-by-ward differences in socio-economic features that can contextualise model estimates. Figure 8 shows model estimates alongside socio-economic characteristics from the synthetic population in the southern edge of the city of Kochi, Kerala. Wards 47, 50, 51, 53, 54 have lower levels of mean estimated LPG consumption than surrounding wards, although model outputs indicate that not all these wards with lower LPG consumption are equal. Amongst these, wards 47, 50 and 53 have a greater proportion of households with residual biomass use (continued biomass use in households with an LPG stove) suggesting that a different intervention may be required to encourage sustained and exclusive use of LPG in these wards. Some local socio-economic context is provided by the proportions of daily and weekly income earners by ward. Notice that many of the wards with high proportions of households paid daily or weekly are also those same wards with higher proportions of residual biomass users, such as wards 47 and 53. Similar outputs are mapped out in Supplementary Figures S.3, S.4, and S.5 for the other cities modelled.

Interestingly, recent studies have shown how multi-level models can help identify and locate residential segregation in cities in the US (Arcaya, Schwartz and Subramanian, 2018) and UK (Jones *et al.*, 2018). While the multi-level model microsimulation approach does not imply causality between variables it does contextualise spatial patterns and inequality in clean cooking access with local socio-economic factors. This can allow stakeholders to not only identify wards in need of further attention, but could provide a starting point for understanding the barriers and features of the transition pathways households in that ward may be following. Such data could also inform the setting of criteria for a particular policy or transition support initiative, or even possibly support the local tailoring of such criteria for national policies.

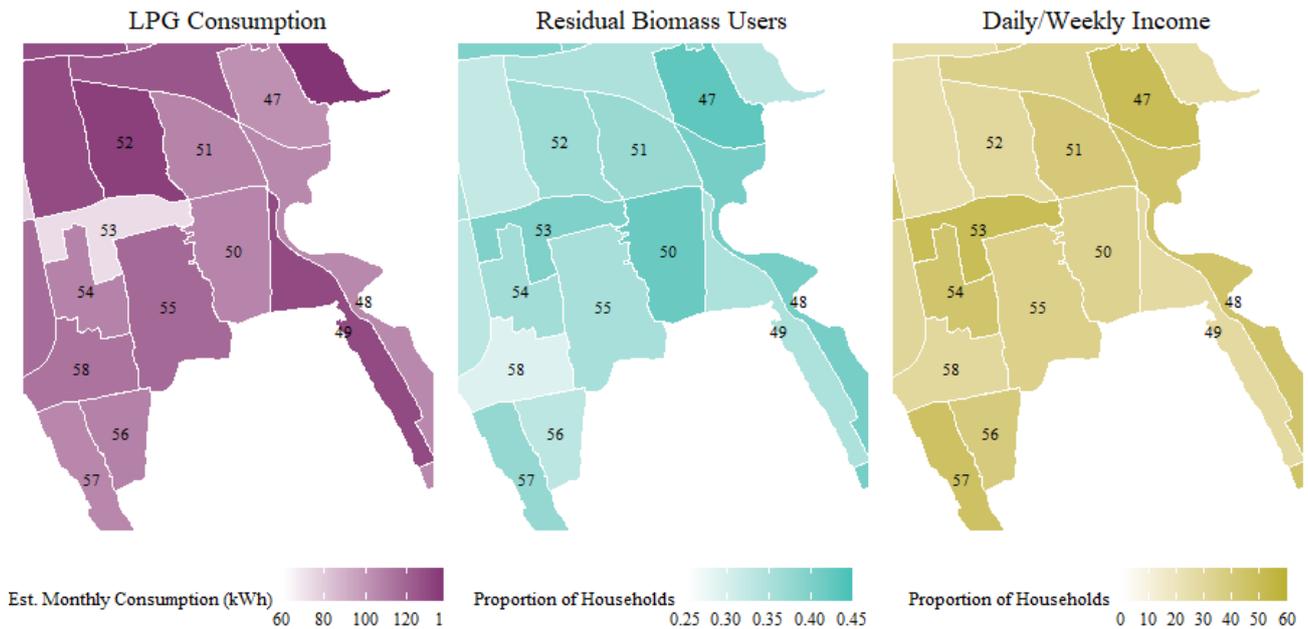

Figure 8: Model outputs for wards in south-eastern section of Kochi, Kerala. Outputs show mean LPG consumption, proportion of households with residual biomass use, and households relying on daily/weekly income.

## Conclusion

A microsimulation approach is proposed to estimate cooking fuel use across a city using publicly available data to help characterise spatial inequality in access to clean cooking. Using a multi-level model which accounts for household cooking practices, local city effects, and spatial effects produced outputs that showed reasonable consistency with real world data. Allowing model parameters to vary by primary cooking fuel group reduced uncertainty in these, while the model structure propagated uncertainty in ward-level effects through to fuel use estimates. There are obviously limitations to this approach some of which relate to the data used and model design. The ward-level aggregation may be too coarse for larger cities where wards have greater population density. Additionally the IHDS and NSS household-level data used may under represent some types of low-income and informal households. Inclusion of different types of data may offer a means to address some of these shortcomings and increase resolution and representation.

The random effect and spatial components capture the 'city effect' and 'ward effect' - the model offers insight into how fuel use in a given city compares to the average urban household in the state, and how the fuel use in a given city ward compares to the average fuel use in the city. By simulating a representative sample of households in each ward, spatial distribution of clean cooking access can be examined in context of socio-economic features across a city. With India's urban population set to expand in the coming decades limited data on the effects and distribution of spatial inequality is a challenge for policy implementation, not only for clean cooking, but wider energy access issues. Design of effective interventions requires stakeholder involvement and consideration of local household needs, and cost-effective modelling approaches such as the one proposed offer an exploratory view of clean cooking access across a city that can facilitate such a process.

## Data availability

All public data sources used can be accessed via the URL in the respective reference. Additionally some survey data was collected for comparison with model outputs. An anonymised version of this data is available at https://doi.org/10.17863/CAM.66449.

## Acknowledgements

A.P. Neto-Bradley is grateful to the EPSRC for support through the CDT in Future Infrastructure and Built Environment (EP/L016095/1). This work was also supported by AI for Science and Government (ASG), UKRI's Strategic Priorities Fund awarded to the Alan Turing Institute, UK (EP/T001569/1) and the Lloyd's Register Foundation programme on Data-centric Engineering. The authors are grateful to Luke Archer at the Leeds Institute for Data Analytics, University of Leeds, for advice on generating synthetic populations using R. The authors would like to thank the Indian Institute for Human Settlements for assistance with shapefiles.

# Supplementary Material

## Study Area

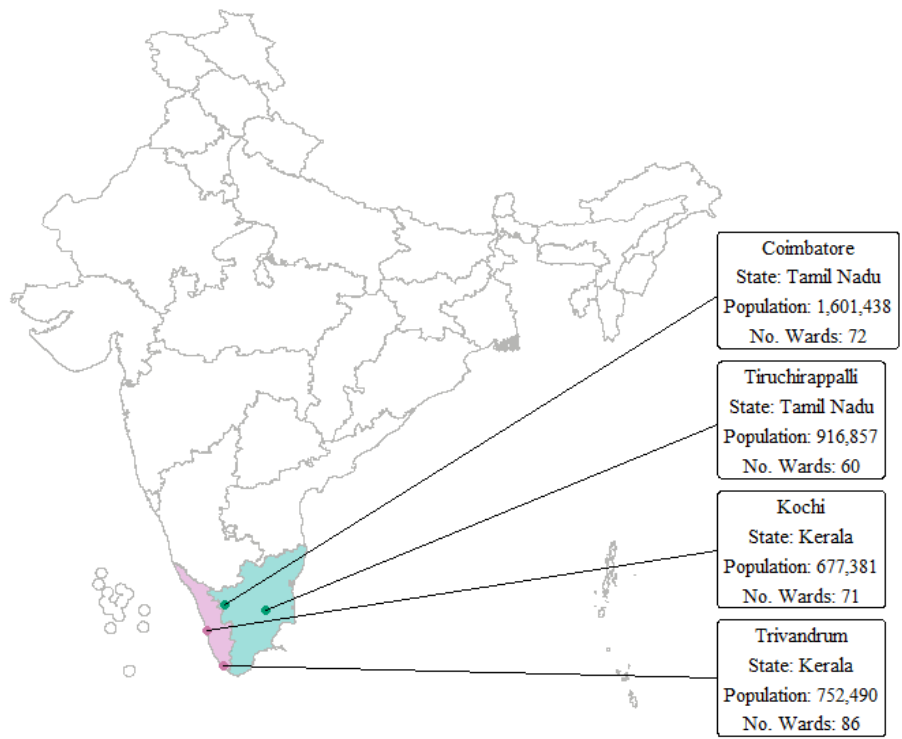

**Figure S.1**: Locations of modelled cities with key demographic characteristics.

This study models the ward level cooking fuel access in four cities located across the southern Indian states of Kerala, and Tamil Nadu. The city locations and population are detailed in Figure S.1. The population of these so-called Tier II cities (not counting wider metropolitan area) sits a little under one million for all except Coimbatore which is slightly more populous than the rest. All cities have larger populations than the average urban area in their states and have seen growth in recent decades. While there are differences between these states in terms of policy and even socio-economic landscape, they all fall within a narrow geographic region with similar climate and geography which ensures a relatively homogeneous physical landscape with minimal differences in effect of climate on energy use, or availability and type of biomass.

## Data

### Model Inputs

The household level data from the Indian Human Development Survey 2011 (IHDS-II) (Desai and Vanneman, 2015) and the 2011 census (Government of India, 2011) ward-level tables of household asset ownership are used to generate the synthetic population. Maps of the Census 2011 ward boundaries, and National Sample Survey consumer survey data from Round 68 (National Sample Survey Organisation, 2013) are used to estimate coefficients of economic determinants in the multi-level model, and spatial effects. While the boundary map must feature the ward divisions of the year of the census, the approach can use the most recently available consumer survey data, which is at a household scale, to provide up to date estimates between the decennial census. Table S.1 details the different scales and variables covered by each dataset. Note that none of these individual datasets contain all the information needed.

**Table S.1:** Scale and variable types available in each of the input datasets for microsimulation.

| Data Source | Scale | Year | Expenditure | Socio-Econ | Fuel Choice | Fuel Use | Geospatial |
|---|---|---|---|---|---|---|---|
| Census | W | 2011 | - | Yes | Yes | - | - |
| IHDS-II | HH | 2011/12 | Yes | Yes | - | Yes | - |
| NSS R68 | HH | 2011/12 | Yes | Yes | Yes | Yes | - |
| GIS Ward Map | W | 2011 | - | - | - | - | Yes |

*Ward-level Survey Data*

To examine the performance of the model we make use of survey data from 24 low-income wards across the four cities. This dataset contains a random sample of 60-70 households within each ward, and includes questions covering demographics, socio-economic factors, and energy consumption and associated behaviours. The survey was conducted between October 2018 and May 2019, and 2011 census ward boundaries were used. Ward selection was on the basis of 2011 Census ward level measures on primary cooking fuel, banking access, and asset ownership with the aim of selecting households scoring poorly on these metrics and thus most likely to exhibit energy access challenges. A simple random sample approach was used and the quantitative basis for sample sizing was to ensure that the estimated mean fuel consumption would have a +/- 10% error at a 95% confidence level, informed by data from previous studies.

## Population Synthesis with Iterative Proportional Fitting

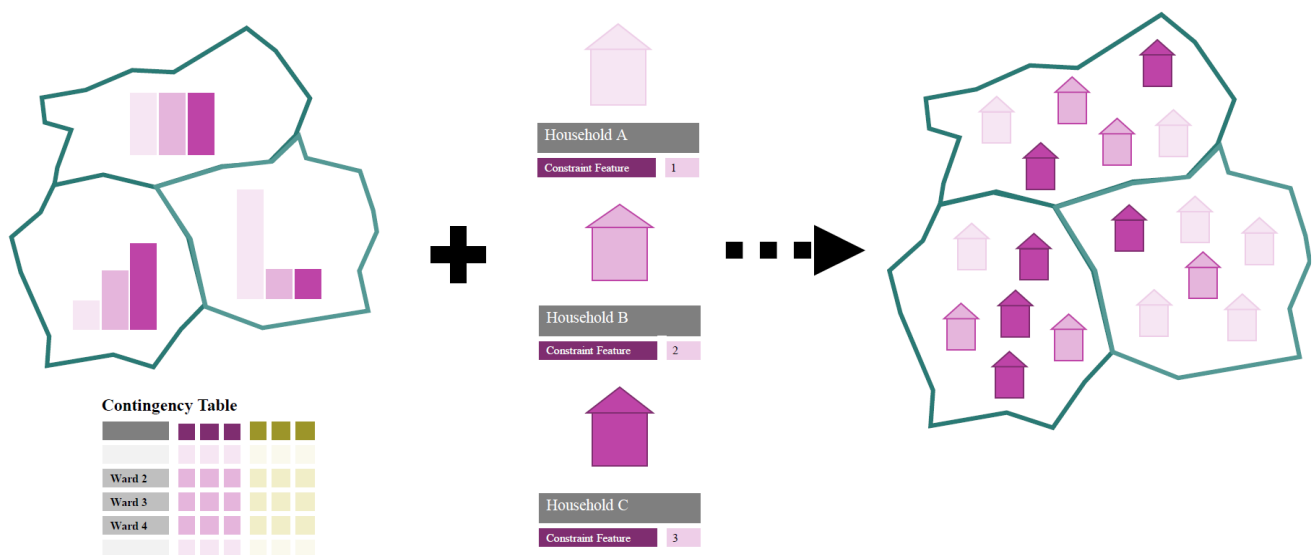

**Figure S.2:** Model outputs for wards in the central-western section of Coimbatore, Tamil Nadu. Outputs show mean LPG consumption, proportion of households with residual biomass use (households with an LPG stove that still use some biomass), and the proportion of households belonging to scheduled castes or tribes.

IPF requires two different types of data, the first being a contingency table of categorical variables for each geographic subdivision, in this case a ward, indicating the totals for each categorical feature within the ward. The census ward level data is used for this and the categories of constraints are detailed in Table S.2. The IHDS is used as the micro data as it contains instances of individuals in the wider area, in this case from a survey of households containing a sample of representative individual households. IPF allocates these individuals from the micro dataset to the zones or, in this case wards, through a weight matrix that indicates how representative each individual is of the specific zone (Lovelace and Dumont, 2016), a process illustrated schematically in Figure S.2.

**Table S.2:** Categories of constraint variables used for IPF contingency table

| Variable | Categories |
| --- | --- |
| Caste | Scheduled Caste, Scheduled Tribe, Other |
| Roof Materials | Thatch, Tile, Stone, Plastic, Metal, Concrete, Brick, Other |
| Wall Material | Thatch, Mud, Plastic, Wood, Brick, Metal, Stone, Concrete, Other |
| Floor Material | Mud, Wood, Brick, Stone, Concrete, Tiles, Other |
| Rooms | 1 to more than 6 |
| Members of household | 1 to more than 6 |
| Home ownership | Rent, Own |
| Primary cooking fuel | Firewood, Crop Residue, Dung, Coal, Kerosene, LPG, None |
| Cooking location | Outside, Inside, No Kitchen |
| Asset Ownership | Bank access, TV, Moped |

IPF will produce fractional weights for each individual in the microdata, and although these would be suitable for producing small area statistics, a motivation for generating a synthetic population is to have actual synthetic individual households with a particular combination of features. Integerisation is the process of essentially rounding these fractional weights and differing approaches to this exist, although it should be noted that integerisation can result in minor discrepancy between the original and synthetic marginal data (Smith et al., 2017). To integerise the fractional weights the approach proposed by Lovelace and Ballas (2013) was used based on truncating, replicating, and sampling which they demonstrated to outperform in accuracy other commonly used methods such as simple rounding, inclusion threshold, counter-weight, and proportional probabilities.

It is important to remember that population synthesis assumes that the micro data (the IHDS household data) is representative of the study area, dependence between variables not constrained for and the constraint variables (those variables present in the census ward level data that are used to determine weightings) is relatively constant, relationships between constraint variables are not spatially dependent, and that the microdata is sufficiently detailed to represent full diversity of the region of study (Lovelace and Dumont, 2016). These assumptions are clearly oversimplifications, in particular that constraint variables are not spatially dependent. This does not invalidate the results, but it is important to understand simplifications made.

Validation of the synthetic population is by its very nature tricky as the synthetic population serves to estimate unknown data (Smith et al., 2009). As a preliminary check, internal validation against the variables in the census contingency table (used to determine weights) ensures that IPF has produced an output faithful to the input data. External validation can be carried out by aggregating features of the synthetic population to a geographic level for which there are known values to compare against (Ballas and Clarke, 2001). Synthetic population data is aggregated at the city level to compare constrained and unconstrained variables to those from the Socio Economic and Caste Census of India from 2011. Checking model estimates against the ward level survey data discussed in the model performance section below also provides a form of external validation on variables which are not used in generating the synthetic population.

# Multi-level Model Variants

## *Basic model: No location based effects*

The basic model starts from an idealised linear relationship for a household's mean fuel use shown in equation S.1. Both household income and household size have been found to be determinants of fuel use (Ahmad and Puppim de Oliveira, 2015; Farsi et al., 2007). For a household $i$, fuel use $\mu_{fuel_i}$ (kWh/month) is a linear combination of household expenditure and household size. Importantly, any given household $i$ belonging to group j may use more than one cooking fuel.

$$\mu_{fuel_i} = a_0 + a_1 exp_i + a_2 size_i \qquad \textbf{Eq. S.1}$$

Coefficients for the household level predictors are allowed to vary by primary cooking fuel group. This is a non-spatial, and non-random group effect (Gelman, 2007). The statistical model for household fuel use is specified in equation S.2, where for a household $i$ in cooking fuel group j fuel use is assumed to follow a normal distribution with a mean given by the relationship in equation S.1, and a precision $\sigma$.

$$fuel_{[i]} \sim N\big(a_{0\,j[i]} + a_{1\,j[i]} exp_{[i]} + a_{2\,j[i]} size_{[i]}, \sigma\big) \qquad \textbf{Eq. S.2}$$

Whose coefficients are defined by parameters as shown:

$$a_{0[i]} \sim N(\mu_{a0}, \sigma_{a0})$$

$$a_{1[i]} \sim N(\mu_{a1}, \sigma_{a1})$$

$$a_{2[i]} \sim N(\mu_{a2}, \sigma_{a2})$$

Parameters for the distributions of coefficients $a_0$, $a_1$, and $a_2$ are estimated from the NSS consumer survey data. A prior is set on the parameters using the fuel use values embedded in the synthetic population, which have been derived from the IHDS micro dataset. Prior mean fuel use (the synthetic household's historical fuel use), $\mu_{fuel_h}$ is assumed to have the same linear relationship with household predictors that is use to estimate current use in equation S.1. A statistical model for $fuel_{h[i]}$ is shown in equation S.3. As the synthetic population does not have corresponding primary fuel choice embedded (this is assigned by categorical logit regression) the prior coefficients do not vary by primary cooking fuel group. Note that for this prior data, $\sigma_h$ is not an estimated parameter as it is a known value and determined from the synthetic population.

$$fuel_{h[i]} \sim N\big(a_{0h[i]} + a_{1h[i]} exp_{h[i]} + a_{2h[i]} size_{h[i]}, \sigma_h\big) \qquad \textbf{Eq. S.3}$$

The coefficients in the linear relationship for mean prior fuel use $\mu_{fuel_h}$, are related to the overall parameters through the following reparameterisation:

$$a_{0h[i]} = \mu_{a0} + \sigma_{a0} z_i$$

$$a_{1h[i]} = \mu_{a1} + \sigma_{a1} z_i$$

$$a_{2h[i]} = \mu_{a2} + \sigma_{a2} z_i$$

$$z \sim N(0,1)$$

Setting a prior using the data embedded in the synthetic population ensures consistency, and provides an initial 'best guess' for coefficients which are then updated using more recent NSS survey state-wide data. Most importantly, this prior has fuel use data for each individual ward and is thus the only data available for estimation of location dependent effects described in the subsequent models. Note that this basic model will account for less uncertainty than the subsequent models given that it does not consider spatial effects or local heterogeneity. The precision terms for the coefficients $\sigma_{a0}$, $\sigma_{a1}$, and $\sigma_{a2}$ represent the uncertainty in the coefficients, while the $\sigma$ in the base model (equation S.2) represents chance variability.

### Random effects model: Non-spatial heterogeneity

This formulation of the model includes an ordinary random effects coefficient to account for non-spatial heterogeneity. As the data used for estimating linear coefficients of $\mu_{fuel\,j[i]}$ is representative of all urban households in the state, the non-spatial heterogeneity captured by this coefficient represents how the given city differs from the average urban area in the state. With respect to the objectives of the model formulation this component captures the influence of local socio-economic context.

The synthetic population's embedded energy use allows for estimation of these ward level coefficients. The formulation of this model in equation S.4 takes the $\mu_{fuel\,j[i]}$ from the basic model (eq. S.2) where coefficients vary by cooking fuel group, with the addition of a random effects coefficient $\theta_{w[i]}$ which varies by the ward a household is in.

$$fuel_{[i]} \sim N\big(\mu_{fuel\,j[i]} + \theta_{w[i]}, \sigma_h\big) \qquad \textbf{Eq. S.4}$$

### ICAR model: Spatial Auto-correlation

This model variant includes an Intrinsic Conditional Auto-Regressive (ICAR) component which accounts for the tendency for adjacent areas to share similar characteristics (Morris et al., 2019). ICAR models are based on an approach developed by Besag (1974), and the advice of Morris et al. (2019) is followed on implementing an ICAR component with Stan. In the ICAR model the spatial component $\phi_w$ is modelled as normally distributed with a mean equal to the average of its neighbours, where the number of neighbours is denoted by $d_w$. The conditional specification is shown in equation S.5, where $w$ is the current ward, $w \sim x$ is the relationship between ward $w$ and its neighbours.

$$p(\phi_w|\phi_{w \sim x}) = N\left(\frac{\Sigma_{w \sim x}\phi_w}{d_w}, \frac{\sigma_w^2}{d_w}\right) \qquad \textbf{Eq. S.5}$$

The ICAR coefficient $\phi$ is estimated from the prior of the synthetic population, and uses a fully connected ward graph as determined from the ward map shapefiles, as an input for the ICAR model. A prior is set on the mean $\phi$ providing a 'soft' sum-to-zero constraint (Morris et al., 2019). This formulation of the fuel use model is shown in equation S.6.

$$fuel_{[i]} \sim N\big(\mu_{fuel\,j[i]} + \phi_{w[i]}, \sigma\big) \qquad \textbf{Eq. S.6}$$

*Combined effects model*

In the combined effect formulation both an ordinary random effects component and a ICAR spatial smoothing component are included along with the linear combination of household level predictors $\mu_{fuel\ j[i]}$ as shown in equation S.7. Again these ward coefficients are estimated using the prior data from the synthetic population, and the ward graph for each city. This model accounts for the most uncertainty given it factors in spatial effects, and local socio-economic and cultural heterogeneity for each ward.

$$\text{fuel}_{[i]} \sim N\big(\mu_{fuel\ j[i]} + \phi_{w[i]} + \theta_{w[i]}, \sigma\big) \qquad \textbf{Eq. S.7}$$

# Additional Model Output Maps

Figure S.3 maps model outputs for a central North-western portion of the city of Coimbatore, Tamil Nadu. This area includes some of the older neighbourhoods and green areas surrounding the agricultural university. There is considerable variance in the spatial distribution of clean cooking uptake. Notice how clusters with higher proportions of lower caste households (e.g. 37, 57, 35, 64) tend to have a higher proportion of household fuel stacking, and low levels of LPG consumption. Adjacent to some of these wards are wealthier neighbourhoods with lower proportions of low caste households such as wards 31, 36, and 27. These have relatively high mean levels of LPG consumption (average of over 100 kWh/month) and below average proportions of households with residual biomass use. Not all wards with greater proportions of low caste households have high residual biomass use though – ward 41 has more average levels of LPG use and residual biomass and this could be a reflection of its central location, with slightly less availability of biomass fuels and proximity to LPG distributors.

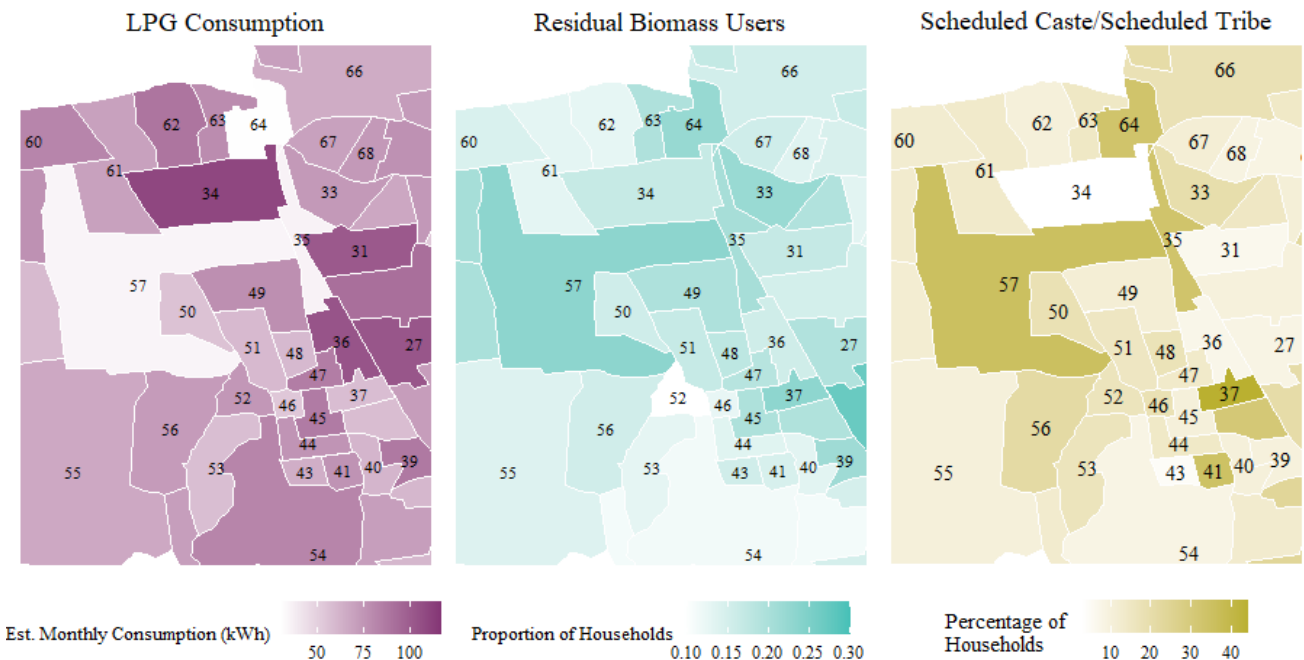

**Figure S.3:** Model outputs for wards in the central-western section of Coimbatore, Tamil Nadu. Outputs show

mean LPG consumption, proportion of households with residual biomass use (households with an LPG stove that still use some biomass), and the proportion of households belonging to scheduled castes or tribes.

Figure S.4 maps model outputs in the central-western part of Trivandrum, Kerala. The wards to the west border the airport (airfield and airport are not counted as a city ward) – notice that these wards (70-85) have lower levels of LPG consumption than those to their immediate East and South. These wards also have noticeably higher levels of households dependent on daily or weekly income. Interestingly, note how the closer the wards are to the airport along the western edge of this section the greater the proportion of household using some residual biomass – this may represent the impact a local availability of firewood/biomass that makes it convenient to continue using some biomass after switching stove (for example for rice cooking which can often be cooked on a biomass stove even when households have LPG), or the undesirability and affordability of the location given proximity to the airport runway.

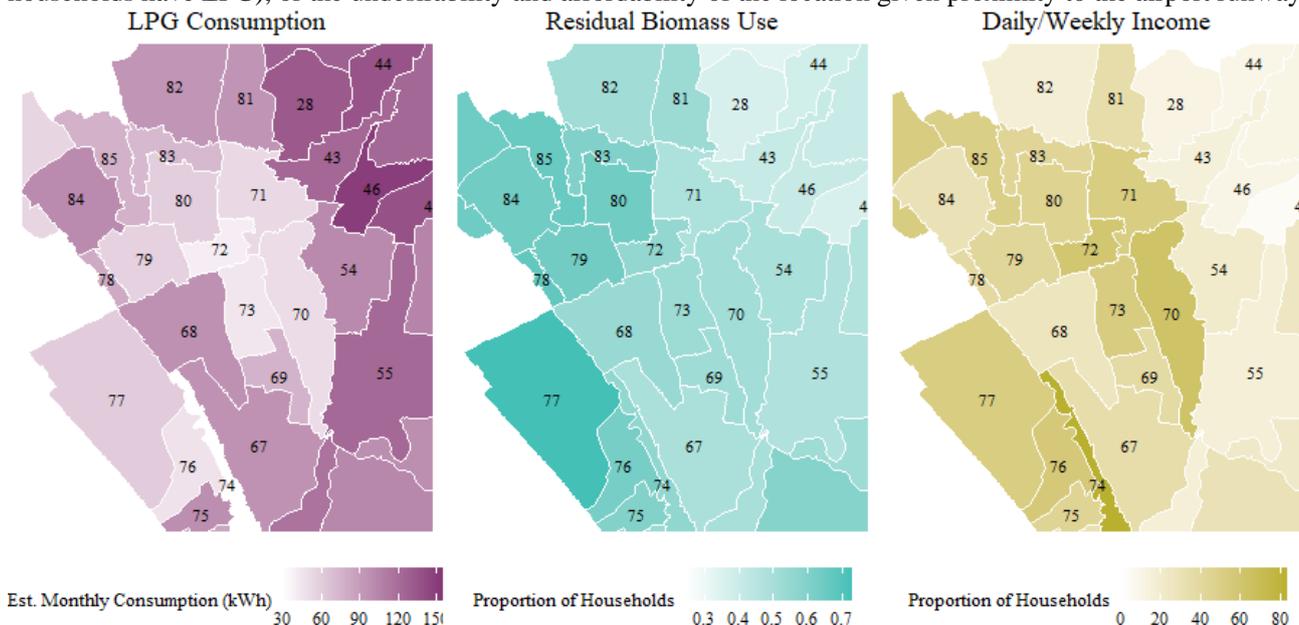

**Figure S.4:** Model outputs for wards in the south-western section of Trivandrum, Kerala. Outputs show mean LPG consumption, proportion of households with residual biomass use (households with an LPG stove that still use some biomass), and the proportion of households in wage labour employment.

Finally, in Figure S.5 shows a section of central Tiruchirappalli in Tamil Nadu which displays some interesting spatial variation in cooking fuel use in the context of access to banking in these wards. Many of these central wards have high levels of LPG use such as 18, 44, 52, 54. However some of these have relatively high levels of residual biomass use – in ward 52 greater availability of such fuels towards the edge of the city may be a factor. The ward with the highest levels of access to formal banking, ward 10, has relatively average levels of LPG consumption in the model but some of the lowest levels of estimated residual biomass use suggesting that these households do not engage in much fuel stacking – it is possible the low LPG consumption indicates use of electricity for cooking in this wealthier ward (recall the model does not estimate magnitude of electricity use).

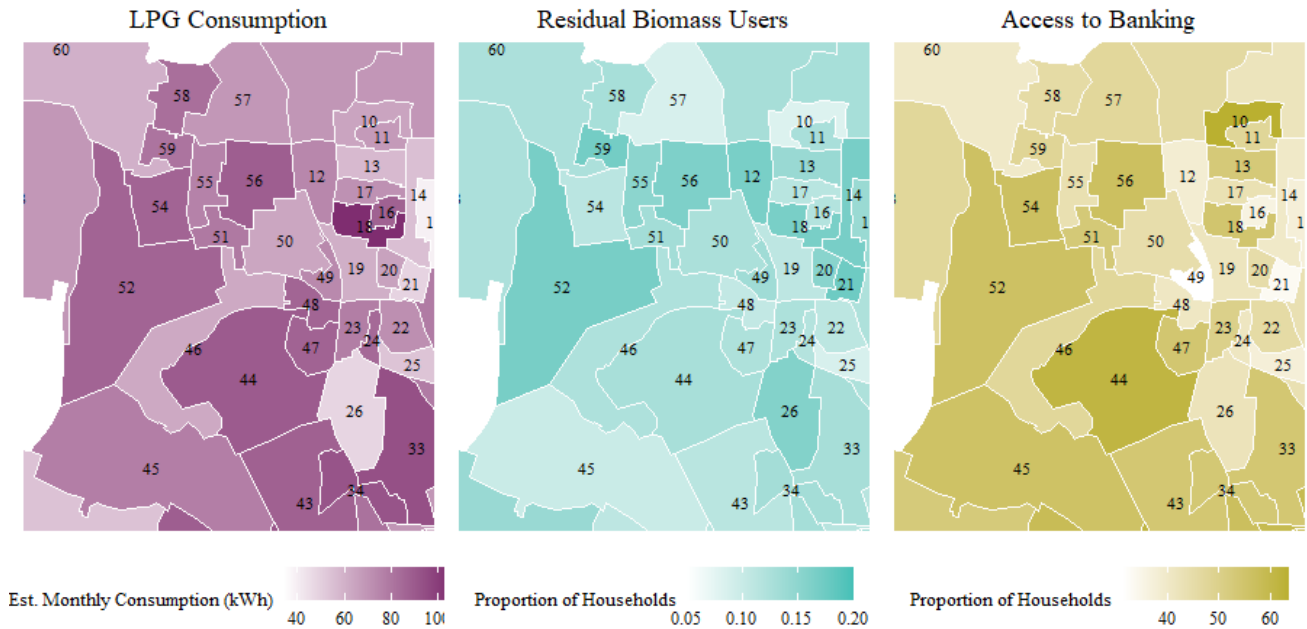

**Figure S.5:** Model outputs for wards in the south-western section of Tiruchirappalli, Tamil Nadu. Outputs show mean LPG consumption, proportion of households with residual biomass use (households with an LPG stove that still use some biomass), and the proportion of households with access to formal banking.